\documentclass[12pt,thmsa]{article}

\usepackage{cite}

\begin{document}

\date{}
\title{Perturbation theory for confined systems}
\author{Francisco M. Fern\'{a}ndez \\
INIFTA (UNLP, CCT La Plata--CONICET), Divisi\'{o}n Qu\'{i}mica Te\'{o}rica \\
Diag. 113 y 64 (S/N), Sucursal 4, Casilla de Correo 16 \\
1900 La Plata, Argentina\\
E--mail:\ fernande@quimica.unlp.edu.ar}
\maketitle

\begin{abstract}
We discuss the application of perturbation theory to a system of particles
confined in a spherical box. A simple argument shows that the particles
behave almost independently in sufficiently strong confinement. We choose
the helium atom with a moving nucleus as a particular example and compare
results of first order with those for the nucleus clamped at the center of
the box. We provide a suitable explanation for some numerical results obtained recently
by other authors.
\end{abstract}

\section{Introduction}

\label{sec:intro}

In a recent paper Montgomery Jr. et al\cite{MAF10} solved the
Schr\"{o}dinger equation for a He atom with its nucleus clamped at the
origin of a box of radius $R_{c}$ with impenetrable walls. They applied
perturbation theory for the case of strong confinement (sufficiently small $%
R_{c}$) and obtained the first five coefficients of the expansion (with
different degrees of accuracy). One of the conclusions in that paper was
that the interaction between the electrons decreases with the box radius.
The authors illustrated this behavior by means of the overlap between the
wavefunctions for the confined He$^{+}$ and for the free electron in the
box.

The purpose of this letter is to discuss those numerical results from a more
general point of view. To this end in Sec.~\ref{sec:PT} we apply
perturbation theory to a system of $N$ particles in a spherical box and
discuss the behaviour of a more general overlap integral. As a particular example, we
compare the energies (corrected through first order) of the He atom when the
nucleus is clamped at the center of the box and when it moves confined in the box.
Finally, in Sec.~\ref{sec:conclusions} we comment on the results and
draw conclusions.

\section{Perturbation theory for strong confinement}

\label{sec:PT}

We consider a system of $N$ particles of masses $m_{i}$ and charges $q_{i}$.
The nonrelativistic Hamiltonian operator is
\begin{equation}
\hat{H}=-\frac{\hbar ^{2}}{2}\sum_{i=1}^{N}\frac{\nabla _{i}^{2}}{m_{i}}%
+\sum_{i=1}^{N-1}\sum_{j=i+1}^{N}\frac{q_{i}q_{j}}{4\pi \epsilon _{0}r_{ij}}
\label{eq:H}
\end{equation}
where $r_{ij}=|\mathbf{r}_{i}-\mathbf{r}_{j}|$ is the distance between the
pair of particles located at $\mathbf{r}_{i}$ and $\mathbf{r}_{j}$.

If the system is confined in a box of radius $R_{c}$ with impenetrable
walls, any solution $\psi $ to the time--independent Schr\"{o}dinger
equation
\begin{equation}
\hat{H}\psi =E\psi   \label{eq:Schro}
\end{equation}
should vanish when $r_{i}\geq R_{c}$ for any given particle $i$. In order to
apply perturbation theory in the case of strong confinement
$R_{c}\rightarrow 0$ we first convert the Schr\"{o}dinger equation (\ref
{eq:Schro}) into a more convenient dimensionless eigenvalue equation. We
choose a representative particle (say $i=1$) and define dimensionless masses
$m_{i}^{\prime }=m_{i}/m_{1}$, charges $q_{i}^{\prime }=q_{i}/q_{1}$ and
coordinates
$\mathbf{r}_{i}^{\prime }=\mathbf{r}%
_{i}/R_{c}$ ($\nabla _{i}^{\prime }=R_{c}\nabla _{i}$).\ We thus obtain a
dimensionless Hamiltonian operator
\begin{eqnarray}
\hat{H}_{d} &=&\frac{m_{1}R_{c}^{2}}{\hbar ^{2}}\hat{H}=-\frac{1}{2}%
\sum_{i=1}^{N}\frac{\nabla _{i}^{\prime 2}}{m_{i}^{\prime }}+\lambda
\sum_{i=1}^{N-1}\sum_{j=i+1}^{N}\frac{q_{i}^{\prime }q_{j}^{\prime }}{%
r_{ij}^{\prime }}  \nonumber \\
\lambda  &=&\frac{R_{c}}{a},\;a=\frac{4\pi \epsilon _{0}\hbar ^{2}}{%
m_{1}q_{1}^{2}}  \label{eq:Hd}
\end{eqnarray}
and the dimensionless eigenvalue equation
\begin{equation}
\hat{H}_{d}\varphi =\epsilon \varphi ,\;\epsilon =\frac{m_{1}R_{c}^{2}}{%
\hbar ^{2}}E=\frac{m_{1}a^{2}\lambda ^{2}}{\hbar ^{2}}E  \label{eq:Schro_dim}
\end{equation}
The new boundary conditions are $\varphi =0$ if any $r_{i}^{\prime }\geq 1$.
Note that if the chosen reference particle is an electron, then $m_{1}=m_{e}$%
, $q_{1}=-e$ and $a=a_{0}$ is the Bohr radius. The transformation just
proposed is a generalization of the one recently applied to the confined
hydrogen atom\cite{F10}.

It is clear that $\hat{H}_{d}(\lambda =0)=\hat{H}_{d}^{0}$ is the
dimensionless Hamiltonian operator for a system of $N$ free particles in a
spherical box of unit radius. Therefore, we can solve the eigenvalue
equation $\hat{H}_{d}^{0}\varphi ^{(0)}=\epsilon ^{(0)}\varphi ^{(0)}$
exactly in terms of products of spherical harmonics and Bessel functions\cite
{F01}. It may also be necessary to consider the permutational symmetry of
the wavefunction and add the corresponding spin factors\cite{MAF10}.

For concreteness, let us consider the He atom. We assume that the particles 1
and 2 are the electrons and the remaining one is the nucleus; that is to
say: $m_{1}=m_{2}=m_{e}$ and $m_{3}=m_{n}$. Obviously, in such a case $%
m_{1}^{\prime }=m_{2}^{\prime }=1$, $m_{3}^{\prime }=m_{n}/m_{e}$ and the
unperturbed wavefunction for the ground state is
\begin{equation}
\varphi ^{(0)}(r_{1}^{\prime },r_{2}^{\prime },r_{3}^{\prime })=2\frac{\sin
(\pi r_{1}^{\prime })}{r_{1}^{\prime }}\frac{\sin (\pi r_{2}^{\prime })}{%
r_{2}^{\prime }}\frac{\sin (\pi r_{3}^{\prime })}{r_{3}^{\prime }}[\alpha
(1)\beta (2)-\beta (1)\alpha (2)]  \label{eq:fi0_He}
\end{equation}
Note that the present model accounts for the motion of the nucleus and that
the  nuclear factor $\sin (\pi r_{3}^{\prime })/r_{3}^{\prime }$ does not
appear if this particle is clamped at the center of the box\cite{MAF10,LC09}.

If we apply straightforward Rayleigh--Schr\"{o}dinger perturbation theory we
obtain the well--known expansions
\begin{equation}
\epsilon =\sum_{j=0}^{\infty }\epsilon ^{(j)}\lambda ^{j},\;\varphi
=\sum_{j=0}^{\infty }\varphi ^{(j)}\lambda ^{j}  \label{eq:PT_series_dim}
\end{equation}
In particular, for the energy we have
\begin{equation}
E=\frac{\hbar ^{2}}{m_{1}a^{2}}\left[ \frac{\epsilon ^{(0)}}{\lambda ^{2}}+%
\frac{\epsilon ^{(1)}}{\lambda }+\epsilon ^{(2)}+\ldots \right]
\label{eq:PT_series}
\end{equation}
that is a generalization of the result derived by Laughlin\cite{L09} and
discussed by Laughlin and Chu\cite{LC09} and Montgomery et al\cite{MAF10}.
Note that equations (\ref{eq:PT_series_dim}) and (\ref{eq:PT_series}) apply
to the most general case of a system of $N$ particles (\ref{eq:Hd}).

If both $\varphi $ and $\varphi ^{(0)}$ are normalized to unity we can
easily prove that
\begin{equation}
\left| \left\langle \varphi \right| \left. \varphi ^{(0)}\right\rangle
\right| \leq 1,\;\lim_{\lambda \rightarrow 0}\left| \left\langle \varphi
\right| \left. \varphi ^{(0)}\right\rangle \right| =1  \label{eq:overlap}
\end{equation}
which clearly account for the behaviour of the overlap integral in Fig.~1 of
Montgomery et al\cite{MAF10} for the particular case of the He$^{+}$. We
stress that Eq.~(\ref{eq:overlap}) applies to the general case of $N$ particles.

As an illustrative example we calculate the energy of the helium atom
corrected through first order.  When the nucleus is clamped at the origin
 we have\cite{MAF10,LC09}
\begin{equation}
\frac{\epsilon (\lambda )}{\lambda ^{2}}=\frac{9.8696044}{\lambda ^{2}}-%
\frac{7.9645404}{\lambda }  \label{eq:E_PT_clamped}
\end{equation}
On the other hand, when the nucleus moves the result is
\begin{equation}
\frac{\epsilon (\lambda )}{\lambda ^{2}}=\frac{9.870280744}{\lambda ^{2}}-%
\frac{5.358219501}{\lambda }  \label{eq:E_PT_moving}
\end{equation}
where we have chosen $m_{n}^{\prime }=7296.300\,m_{e}$. It is worth noting
that the effect of the nuclear motion is more noticeable on
the average potential energy than on the kinetic energy.\cite{F01}.

\section{Conclusions}

\label{sec:conclusions}

We have shown that converting the Schr\"{o}%
dinger equation into a dimensionless eigenvalue one greatly facilitates the
application of perturbation theory to strongly confined systems. In
particular this approach clearly shows that the interaction between the
particles becomes negligible as the confinement increases. In this way we
could provide a suitable mathematical basis for recent numerical
calculations on the He atom with a nucleus clamped at origin. Eq.~(\ref
{eq:overlap}) not only explains the behavior of the overlap integral
calculated by Montgomery et al\cite{MAF10} but also reveals that the same
kind of curve should be expected for any system of particles confined in a
spherical box.

We have also shown that the effect of the nuclear motion on the kinetic
energy of the confined atom is not as important as its effect on the average
of the Coulomb interactions. The reason is that the energy of the confined
atom changes markedly with the location of the clamped nucleus. Therefore,
when it moves there is a sort of average contribution to the potential
energy from all the possible locations inside the box.

\end{document}